\documentclass{emulateapj}

\newcommand{\numax}{\mbox{$\nu_{\rm max}$}}
\newcommand{\Dnu}{\mbox{$\Delta \nu$}}
\newcommand{\dnu}[1]{\mbox{$\delta \nu_{#1}$}}
\newcommand{\muHz}{\mbox{$\mu$Hz}}
\newcommand{\kep}{\mbox{\textit{Kepler}}}

\slugcomment{Accepted for publication in the Astrophysical Journal}

\shorttitle{Global oscillation parameters of red giants observed by \textit{Kepler}}
\shortauthors{D. Huber et al.}

\begin{document}

\title{Asteroseismology of red giants from the first four months of \textit{KEPLER} data: 
Global oscillation parameters for 800 stars}

\author{
D.~Huber\altaffilmark{1}, 
T.~R.~Bedding\altaffilmark{1}, 
D.~Stello\altaffilmark{1}, 
B.~Mosser\altaffilmark{2}, 
S.~Mathur\altaffilmark{3}, 
T.~Kallinger\altaffilmark{4,5}, 
S.~Hekker\altaffilmark{6}, 
Y.~P.~Elsworth\altaffilmark{6}, 
D.~L.~Buzasi\altaffilmark{7}, 
J.~De~Ridder\altaffilmark{8}, 
R.~L.~Gilliland\altaffilmark{9}, 
H.~Kjeldsen\altaffilmark{10}, 
W.~J.~Chaplin\altaffilmark{6}, 
R.~A.~Garc\'\i a\altaffilmark{11}, 
S.~J.~Hale\altaffilmark{6}, 
H.~L.~Preston\altaffilmark{7,12}, 
T.~R.~White\altaffilmark{1}, 
W.~J.~Borucki\altaffilmark{13}, 
J.~Christensen-Dalsgaard\altaffilmark{10}, 
B.~D.~Clarke\altaffilmark{14}, 
J.~M.~Jenkins\altaffilmark{14}, and 
D.~Koch\altaffilmark{13} 
}
\altaffiltext{1}{Sydney Institute for Astronomy (SIfA), School of Physics, University of Sydney, NSW 2006, Australia; \mbox{dhuber@physics.usyd.edu.au}}
\altaffiltext{2}{LESIA, CNRS, Universit\'e Pierre et Marie Curie, Universit\'e Denis, Diderot, Observatoire de Paris, 92195 Meudon cedex, France}
\altaffiltext{3}{High Altitude Observatory, NCAR, P.O. BOX 3000, Boulder, CO 80307, USA}
\altaffiltext{4}{Department of Physics and Astronomy, University of British Columbia, Vancouver, Canada}
\altaffiltext{5}{Institute for Astronomy, University of Vienna, 1180 Vienna, Austria}
\altaffiltext{6}{School of Physics and Astronomy, University of Birmingham, Edgbaston, Birmingham B15 2TT, UK}
\altaffiltext{7}{Eureka Scientific, 2452 Delmer Street Suite 100, Oakland, CA 94602-3017, USA}
\altaffiltext{8}{Instituut voor Sterrenkunde, K.U.Leuven, Belgium}
\altaffiltext{9}{Space Telescope Science Institute, 3700 San Martin Drive, Baltimore, Maryland 21218, USA}
\altaffiltext{10}{Danish AsteroSeismology Centre (DASC), Department of Physics and Astronomy, Aarhus University, DK-8000 Aarhus C, Denmark}
\altaffiltext{11}{Laboratoire AIM, CEA/DSM-CNRS, Universit\'e Paris 7 Diderot, IRFU/SAp, Centre de Saclay, 91191, Gif-sur-Yvette, France}
\altaffiltext{12}{Department of Mathematical Sciences, University of South Africa, Box 392 UNISA 0003, South Africa}
\altaffiltext{13}{NASA Ames Research Center, MS 244-30, Moffett Field, CA 94035, USA}
\altaffiltext{14}{SETI Institute, NASA Ames Research Center, MS 244-30, Moffett Field, CA 94035, USA}

\begin{abstract}
We have studied solar-like oscillations in $\sim$\,800 red-giant stars using \kep\ long-cadence 
photometry. The sample includes stars ranging in evolution from the lower part of the red-giant 
branch to the Helium main sequence. We investigate the relation between the large frequency 
separation (\Dnu) and the frequency of maximum  power (\numax) and show that it is 
different for red giants than for main-sequence stars, which is consistent with evolutionary 
models and scaling relations. The distributions of \numax\ and \Dnu\ are in qualitative agreement 
with a simple stellar population model of the Kepler field, including the first evidence for a secondary 
clump population characterized by $M\gtrsim2\,M_{\sun}$ and $\numax\simeq40-110\,\muHz$.  We measured 
the small frequency separations \dnu{02} and \dnu{01} in over 400 stars and  \dnu{03} in over 
40. We present C-D diagrams for $l=1$, 2 and 3 and show that the frequency  separation ratios 
\dnu{02}/\Dnu\ and \dnu{01}/\Dnu\ have opposite trends as a function of \Dnu. The data show a 
narrowing of the $l=1$ ridge towards lower \numax, in agreement with models predicting more  
efficient mode trapping in stars with higher luminosity. We investigate the offset $\epsilon$ in 
the asymptotic relation and find a clear correlation with \Dnu, demonstrating that it is related to 
fundamental stellar parameters. Finally, we present the first amplitude-\numax\ relation 
for \kep\ red giants. We observe a lack of low-amplitude stars for $\numax\gtrsim110\,\muHz$ and 
find that, for a given \numax\ between $40-110\,\muHz$, stars with lower \Dnu\ (and consequently 
higher mass) tend to show lower amplitudes than stars with higher \Dnu.
\end{abstract}


\section{Introduction}

Stars with convective envelopes show solar-like oscillations that are 
sensitive to the physical processes governing their interiors \citep[see, e.g.,][]{B+G94, ChD2004}. 
Following the success of helioseismology, the detection of such oscillations in a variety of stars 
holds great promise for improving our understanding of stellar structure and evolution. 

The traditional goal of asteroseismology is the accurate measurement of individual mode frequencies, 
which can be used to test stellar physics by comparing them to 
frequencies predicted by models. In a more general approach, global oscillation parameters can be 
used. These include the average frequency separations, which
are directly related to properties of the sound speed in the stellar 
interior and therefore to fundamental stellar parameters.
Other parameters are the amplitude and central frequency of the oscillation envelope, which 
are important for understanding the physics of driving and damping of these modes.
The measurement of these parameters presents a valuable addition to classical methods such as 
spectroscopy, and a powerful tool to systematically study stellar evolution when oscillations are 
detected in a large ensemble of stars. 

Compared to main-sequence stars, red giants pulsate with larger amplitudes and longer periods, 
therefore requiring less sensitivity but longer and preferably continuous time series for unambiguous 
detections. The first attempts to detect oscillations in G and K giants were focused on nearby targets such as Arcturus 
\citep{smith83,cochran88,belmonte90}. This was followed by campaigns targeting 
single stars and clusters using precise ground-based Doppler spectroscopy 
\citep{Frandsen02,deridder06} and photometry \citep{stello06}, 
as well as using space-based photometers such as the \textit{HST} \citep{edmonds96,gilliland08,stello09b}, 
\textit{WIRE} \citep{buzasi,retter03,stello08}, \textit{MOST} \citep{barban07,kallinger08b,kallinger08} 
and \textit{SMEI} \citep{tarrant07}. 

A breakthrough was achieved by \textit{CoRoT} 
with unambiguous detections of radial and non-radial modes in $\sim$\,800 red 
giants \citep{deridder09,hekker09,carrier10}. With a maximum time series length of 150\,d but limitations 
due to the low-Earth orbit of the satellite, the \textit{CoRoT} detections were focused on 
low-mass He-core burning stars (the red clump). Using these data, \citet{miglio09} performed the first 
population study using global oscillation parameters and concluded that the distributions were qualitatively
in agreement with the current picture of the star formation rate in our galaxy. \citet{mosser10} 
subsequently used a larger number of detections in the \textit{CoRoT} sample to investigate correlations of 
various oscillation parameters. 

A new era of ``ensemble asteroseismology'' was recently entered 
with the launch of the \kep\ space telescope \citep{gilliland10}. First results for red giants were
based on 34\,d of data 
\citep[Bedding et al. 2010b; see also][]{stello10,hekker10} which demonstrated the enormous potential of \kep\ data. 
While \citet{BHS10} focused on a sample of low-luminosity giants, the goal of this paper is to 
systematically investigate global oscillation parameters in the complete \kep\ red-giant sample 
using data spanning up to 138\,d. We refer to our companion papers for 
the comparison of global oscillation parameters derived using different methods 
\citep{hekker_comp} and the asteroseismic determination of stellar masses and radii \citep{kall_comp}.

\section{Observations and data analysis}

The \kep\ space telescope was launched in March 2009 with the principal science goal of detecting 
Earth-like planets around solar-like stars through the observation of photometric transits. \kep\ 
employs two observation modes, sampling data either in 1\,min (short-cadence) or 29.4\,min 
(long-cadence) intervals. For our study of pulsations in red giants we used \kep\ long-cadence data 
\citep{jenkins10}, which have a Nyquist frequency of 283\,\muHz.

\kep\ is located in an Earth-trailing orbit with spacecraft rolls performed at quarterly intervals 
to redirect solar panels towards the Sun. Data are consequently subdivided into quarters, starting with 
the initial commissioning run (10\,d, Q0), followed by a short first quarter (34\,d, Q1) and 
a full second quarter (90\,d, Q2). The basis of our study consists of 1531 light curves for 
which data from all these quarters are available (total time span of 137.9\,d). We did not 
include stars thought to be members of the open clusters in the \kep\ field (NGC6791, NGC6811, NGC6819 
and NGC6866).

Before we extracted asteroseismic information from the time series, two main instrumental 
artifacts had to be addressed. Firstly, in some cases data from different quarters show intensity 
discontinuities that are mainly caused by pixel shifts after reorientation of the spacecraft \citep{jenkins10}. 
Additionally, two safe-mode events in Q2 caused intensity drifts due to thermal effects, 
affecting in total $\sim$\,7.5 days of the data. We discarded data affected by the safe-mode events 
and corrected for 
intensity jumps by robust fitting and correcting second-order polynomials to Q0, Q1 and five separate 
segments of Q2 which were unaffected by intensity jumps \citep[see also][]{kall_comp}. 
Finally, we performed a simple 4-$\sigma$ outlier clipping to remove remaining outlying data points 
(in most cases $<$\,0.1\% of the total number of data points). 

\begin{figure}
\begin{center}
\resizebox{\hsize}{!}{\includegraphics{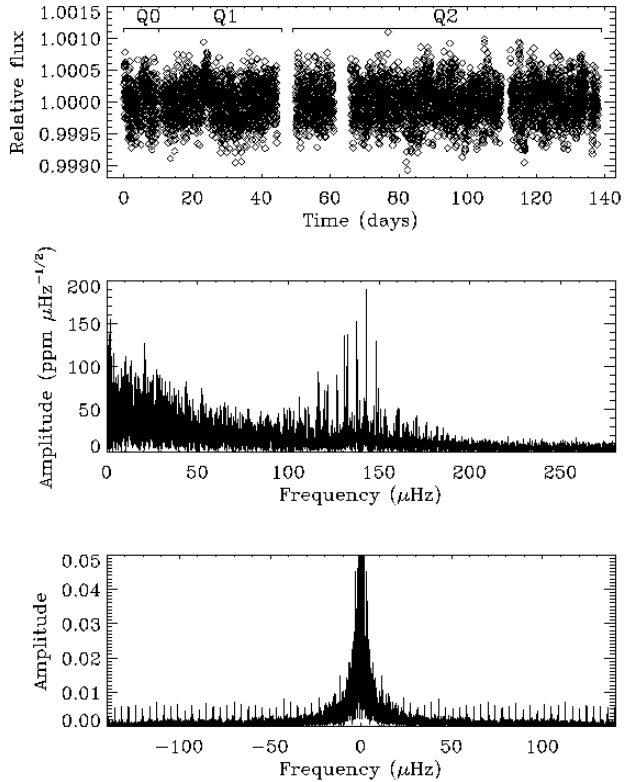}}
\caption{\textit{Top panel:} \kep\ light curve of the oscillating red giant KIC5515314. The different quarters 
of data used in this work are indicated. \textit{Middle panel:} Amplitude spectrum of the light curve 
shown in the top panel. Note that the ordinate shows the square root of power density,  where 
the latter is power multiplied by the effective length of the dataset (calculated as the inverse of the area under the 
spectral window). \textit{Bottom panel:} Spectral window, shown at the same abscissa scale 
as the middle panel. Note that the height of the peak at zero frequency is 1.}
\label{fig:lc}
\end{center}
\end{figure}

The top panel of Figure \ref{fig:lc} shows a typical light curve after the 
above corrections have been performed. The different quarters contributing to the final 
light curve are indicated. The main gaps in the datasets span 1.5\,d between Q0 and Q1, 4.5\,d 
between Q1 and Q2, and 5\,d as well as 2.5\,d during the aforementioned safe-mode events in Q2. 
The middle panel displays the amplitude spectrum of the light curve, showing the clear 
signature of solar-like oscillations with regularly spaced peaks centered around 130\,\muHz. The 
spectral window shown in the bottom panel of Figure \ref{fig:lc} demonstrates the nearly continuous sampling of 
\kep\ data. The weak peaks at multiples of 3.96\,\muHz\ are caused by the angular momentum 
dumping cycle of the spacecraft causing one rejected datapoint every 2.9\,d. At a level of less than 
1\% in amplitude, these aliasing artifacts are negligible for the analysis considered 
in this paper. 

To extract oscillation parameters from the corrected time series, several methods 
have been employed \citep{hekker09b,HSB09,kall_comp,mathur10,mosser09}. For a detailed comparison 
of the results of these different methods we refer to \citet{hekker_comp}. Unless otherwise mentioned, 
all results shown here are based on the method by \citet{HSB09}. All values from that method 
were cross-checked with the 
others and we only retained stars for which at least one other method yielded 
consistent results. Uncertainties reported in this paper are based on calibrations of
extensive ($\sim$\,50000) simulations of artificial time series with identical sampling to the 
\kep\ data, including mixed modes, background noise and varying mode lifetimes 
between 15--75\,d.

Before reporting our results, we provide a brief summary of the global oscillation parameters used 
in this work, their physical interpretation and the principal methods with which they were determined.

\section{Global oscillation parameters}

\subsection{Frequency of maximum power (\numax)}
 
As first argued by \citet{brown91}, \numax\ for sun-like stars is expected to scale with the 
acoustic cut-off frequency and can therefore be related to fundamental 
stellar parameters, as follows \citep{kb95}:

\begin{equation}
\nu_{\rm max} = \frac{ M/M_{\sun}(T_{\rm eff}/T_{\rm eff,\sun})^{3.5}}{L/L_{\sun}} \nu_{\rm max,\sun} \: .
\label{equ:nmax}
\end{equation}

\noindent
As shown by \citet{stello09} for stellar models and as observed in stars with well-determined 
fundamental parameters \citep[see, e.g.,][]{stello08}, this scaling relation also holds for red giants.
In these stars, \numax\ is a good indicator of the evolutionary stage, and ranges from $\sim$\,10\,\muHz\ 
for high-luminosity giants to $\sim$\,250\,\muHz\ for H-shell-burning stars in the lower part of the red-giant 
branch. It is measured by determining the maximum of the power excess after heavily smoothing 
over several orders or by fitting a Gaussian function to the excess power.

\begin{figure}
\begin{center}
\resizebox{\hsize}{!}{\includegraphics{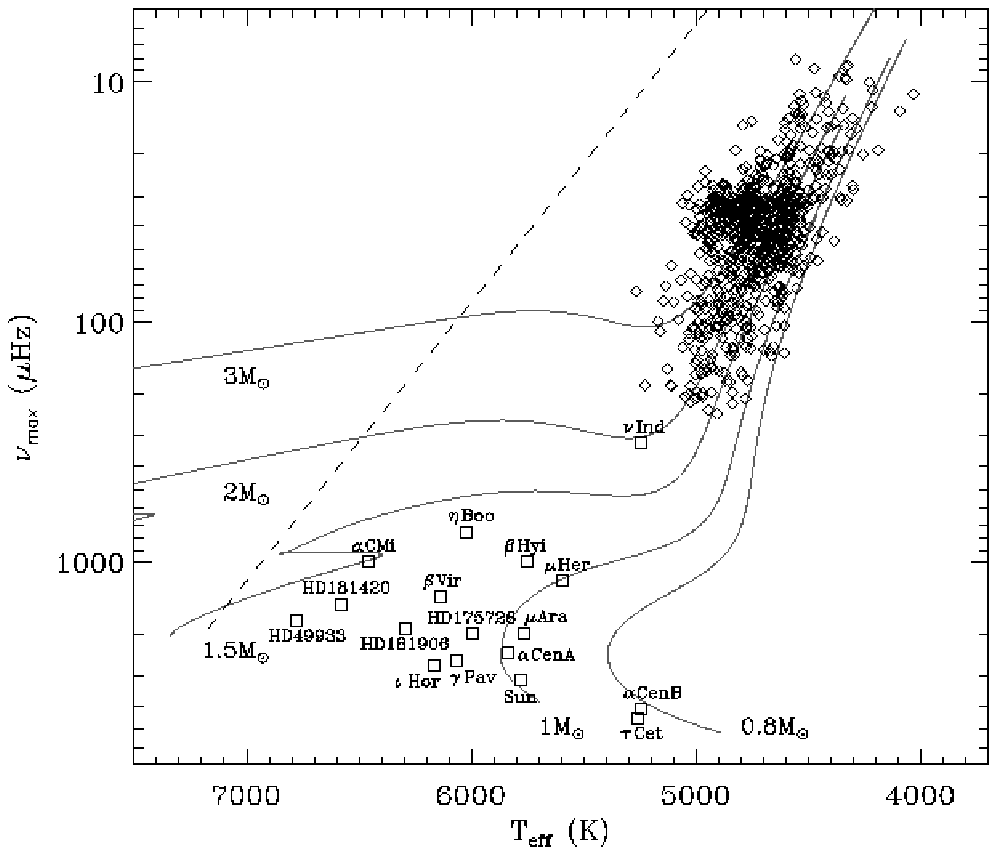}}
\caption{\numax\ versus effective temperature for all red giants in our sample (diamonds), 
as well as main-sequence and sub-giant stars studied by \citet{stello09} (squares). Error bars 
have been omitted for clarity. The grey lines show solar-metallicity {\sc{ASTEC}} 
evolutionary tracks \citep{astec} for a range of masses (note that the 0.8\,$M_{\sun}$ track has been 
evolved beyond the age of the universe). The dashed line marks the approximate position of 
the red edge of the instability strip.}
\label{fig:hrd}
\end{center}
\end{figure}

To illustrate our sample, Figure \ref{fig:hrd} shows all stars in a plot of \numax\ versus $T_{\rm eff}$. 
Effective temperatures have been taken from the Kepler Input Catalog (KIC) 
\citep{latham05} and we have overlaid solar-scaled {\sc{ASTEC}} evolutionary tracks \citep{astec}. 
In addition to the red giants, we also show a sample of main-sequence and sub-giant stars 
for which oscillations had been detected prior to the \kep\ mission 
\citep[see][and references therein]{stello09}.
Figure \ref{fig:hrd} can be viewed as an asteroseismic H-R diagram in which, in the absence of 
parallaxes, we have used 
$1/\nu_{\rm max}$ instead of luminosity. As can be seen from the evolutionary tracks in 
Figure \ref{fig:hrd}, our sample 
spans a total range of masses of approximately 1--3\,$M_{\sun}$, with temperatures ranging from 4200 to 
5200\,K. For a detailed study of the fundamental parameters of the stars in the sample we refer to our 
companion paper \citep{kall_comp}.

\subsection{Frequency separations}
\label{sec:spacings}

According to the asymptotic relation for modes of low angular degree $l$ and high radial order $n$ 
\citep{Tas80,gough86}, frequencies of solar-like oscillations can be described by a series of 
characteristic separations. Observationally, these separations can be expressed as follows 
\citep{bedding_scaling}:

\begin{equation}
\nu_{n,l} = \Delta\nu(n + \frac{1}{2}l + \epsilon) - \delta\nu_{0l} \: .
\label{equ:asymt}
\end{equation}

\noindent
Here, \Dnu\ denotes the mean large frequency separation of modes with the same degree and 
consecutive order. 
\Dnu\ is directly related to the sound travel time across the stellar diameter 
and probes the mean stellar density \citep{ulrich86}. This means that \Dnu\ is expected to scale as 
follows:

\begin{equation}
\Delta\nu = \frac{(M/M_{\sun})^{0.5}(T_{\rm eff}/T_{\rm eff,\sun})^{3}}{(L/L_{\sun})^{0.75}} \Delta\nu_{\sun} \: .
\label{equ:dnu}
\end{equation}

\noindent
In Equation (\ref{equ:asymt}), \dnu{0l}\ denotes the small frequency separations of non-radial modes relative 
to radial modes, as follows:

\begin{equation}
\delta\nu_{02} = \nu_{n,0} - \nu_{n-1,2} \: ,
\label{equ:d02}
\end{equation}
\begin{equation}
\delta\nu_{01} = \frac{1}{2} (\nu_{n,0} + \nu_{n+1,0}) - \nu_{n,1} \: ,
\label{equ:d01}
\end{equation}
\begin{equation}
\delta\nu_{03} = \frac{1}{2} (\nu_{n,0} + \nu_{n+1,0}) - \nu_{n,3} \: .
\label{equ:d03}
\end{equation}

\noindent
Following \citet{BHS10}, we have used \dnu{03} instead of the more commonly used $\delta\nu_{13}$ due 
to the broadening of the $l=1$ ridge in red giants caused by mixed modes 
\citep{dupret,deheuvels}. For main-sequence stars, small frequency separations are sensitive to 
variations of the sound speed gradient near the stellar core, which changes as the star evolves due to the 
increase in its mean molecular weight. 

The phase constant $\epsilon$ in Equation (\ref{equ:asymt}) has contributions from the inner and outer 
turning point of the modes. This is usually expressed as $\epsilon = \frac{1}{4} + \alpha$, where 
$\alpha$ is the contribution from the outer turning point, which is determined by the properties of 
the near-surface region of the star \citep{epsilon}.

\subsection{Maximum mode amplitude}
\label{sec:amplitudes}

The maximum mode amplitude is related to the turbulent convection mechanisms that excite and damp the 
oscillations. \citet{kb95} found that model predictions of solar-like oscillations by \citet{CDF83} 
implied a scaling for velocity amplitudes of 

\begin{equation}
v_{\rm osc} \propto \left(\frac{L}{M} \right)^{s} \: ,
\label{equ:amps}
\end{equation}

\noindent
with $s=1.0$ \citep[see also][]{HBChD99}. More recent calculations for main-sequence stars by 
\citet{SGT2007} have favored an exponent of $s=0.7$.

\citet{kb95} also argued that the oscillation amplitude $A_{\lambda}$ observed in photometry at wavelength 
$\lambda$ is related to the velocity amplitude:

\begin{equation}
A_{\lambda} \propto \frac{v_{\rm osc}}{\lambda} T_{\rm eff}^{-r} \: ,
\label{equ:amps2}
\end{equation}

\noindent
where $r=1.5$ if the oscillations are adiabatic. However, \citet{kb95} found a better fit 
to observed amplitudes in classical pulsators with an exponent of \makebox{$r=2.0$}.

Observationally, \citet{stello10} found agreement with $s=0.7$ (assuming $r=2.0$) for cluster red giants 
observed with \kep, while \citet{mosser10} found a best fitting value of $s=0.89\pm 0.02$ 
(assuming $r=1.5$) for \textit{CoRoT} red giants. As for \numax, mode amplitudes are usually determined by 
heavily smoothing the power spectrum to eliminate variations caused by the stochastic nature of the signal 
\citep{KB08} or by fitting a Gaussian function to the power excess envelope. 
In both cases, it is important to determine an accurate fit to the stellar background contribution. 
To account for the effects of averaging from the 29.4\,min integrations, we 
divided the observed amplitude with a sinc function:
$A_{\rm real} = A_{\rm obs}/ \rm sinc(\frac{\pi}{2} \frac{\nu_{\rm max}}{\nu_{\rm Nyq}})$.

\section{Results}

\subsection{The \Dnu\ - \numax\ relation}
The growing number of detections of solar-like oscillations across the HR diagram has revealed 
a tight power-law relation between \Dnu\ and \numax:

\begin{figure}
\begin{center}
\resizebox{\hsize}{!}{\includegraphics{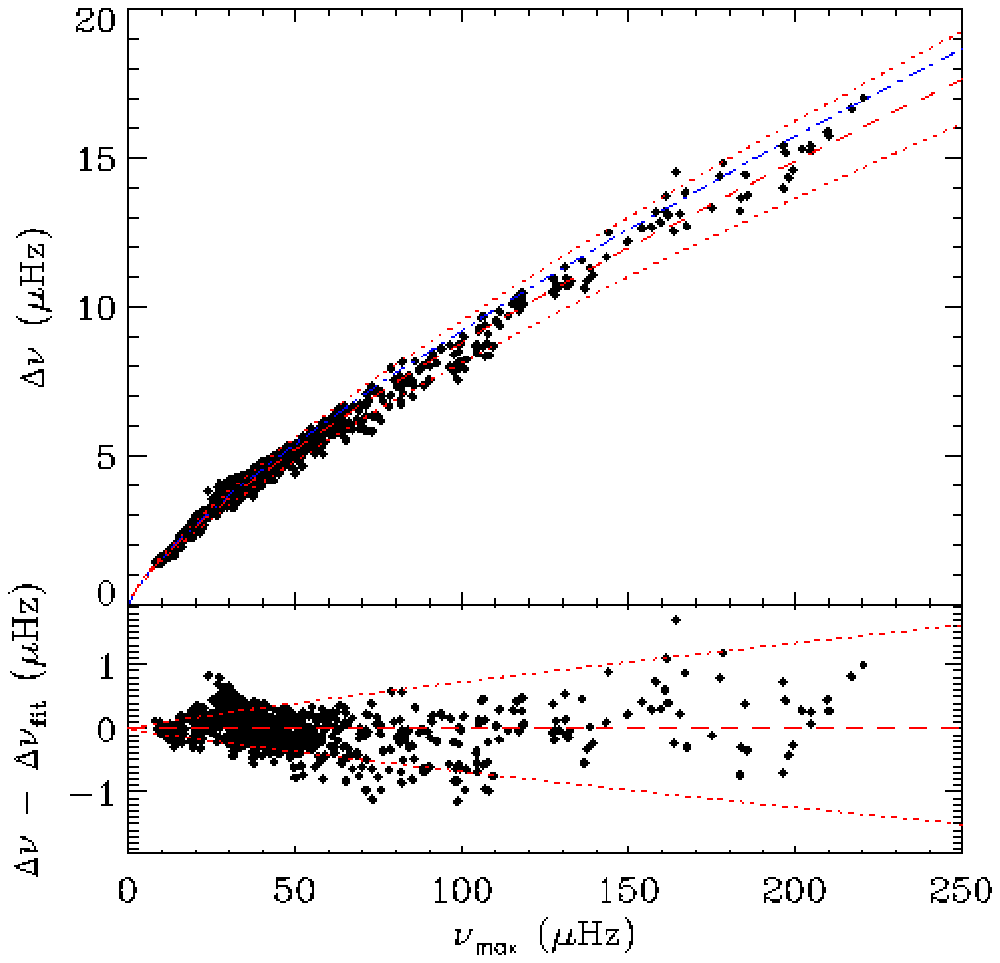}}
\caption{\textit{Upper panel:} \Dnu\ versus \numax\ for the \kep\ red giants. 
The fitted relation is indicated by the red dashed line, with red dotted lines marking 3-$\sigma$ uncertainties. 
The blue dashed-dotted line shows the relation by \citet{stello09}.
Note that the relation by \citet{mosser10}  and the average relation for \kep\ stars derived using 
seven different analysis methods are indistinguishable from the red dashed line. 
\textit{Lower panel:} Residuals after subtracting the fit indicated by the red dashed line.}
\label{fig:02}
\end{center}
\end{figure}

\begin{equation}
\Delta\nu = \alpha (\nu_{\rm max}/\mu\rm Hz)^\beta \: .
\label{equ:dnunumax}
\end{equation}

\noindent
Using a fit to red giants and main-sequence stars, \citet{stello09} derived $\alpha=0.263\pm 0.009\,\muHz$ 
and $\beta=0.772 \pm 0.005$, noting that models predicted slightly different relations for 
red giants than for main-sequence stars (see their Figure 3).  The relation was also seen for the 
\textit{CoRoT} red giants ($\nu_{\rm max}<100\,\muHz$) by \citet{hekker09} and subsequently by
\citet{mosser10}, with the latter finding $\alpha=0.28\pm0.02\,\muHz$ and $\beta=0.75 \pm 0.01$.

The \kep\ data allow us to investigate this relation for the first time over a large part of the red 
giant branch. Figure \ref{fig:02} shows the relation for our sample. 
To derive the coefficients in Equation (\ref{equ:dnunumax}), we performed 
least-squares fits to the results of seven different methods. Note that we have omitted all stars with 
$\numax>230\,\muHz$ from the analysis of the \Dnu-\numax\ relation due to the difficulty of determining 
accurate \numax\ values close to the Nyquist frequency. As can be seen in Table 1,
there is good agreement between the results of different methods, as well as with the result 
by \citet{mosser10} that was based on \textit{CoRoT} stars with $\numax<100\,\muHz$. 

As already shown by \citet{stello09}, a value of $\beta\sim 0.75$
is expected from the scaling relations for \numax\ and \Dnu. Raising Equation 
(\ref{equ:nmax}) to the power of 0.75 and dividing by Equation (\ref{equ:dnu}) yields 

\begin{equation}
\frac{(\numax/\muHz)^{0.75}}{\Dnu/\muHz} \propto \left(\frac{M}{M_{\sun}}\right)^{0.25} \left(\frac{T_{\rm eff}}{T_{\rm eff,\sun}}\right)^{-0.375} \: .
\label{equ:dnunumax2}
\end{equation}

\noindent
This removes the dependence on luminosity and leaves only weak dependences on
mass and effective temperature, with the latter only varying by a small amount on the red-giant branch.

\begin{table} 
\begin{center}
\caption{Coefficients of the \Dnu-\numax\ relation.}
\begin{tabular}{l c c c c}        
\hline         
\hline
Source			&	$\alpha$		&	$\beta$					& \numax		& \# of stars	\\ 
				&	(\muHz)			&							& (\muHz)	&				\\
\hline
A2Z$^{1}$		& 0.266$\pm$0.004	&	0.759$\pm$0.003			&	8--221		& 810		\\			
CAN$^{1}$		& 0.286$\pm$0.004	&	0.745$\pm$0.003 		&	12--228		& 893		\\
COR$^{1}$		& 0.279$\pm$0.004	&	0.749$\pm$0.003 		&	2--208		& 1151		\\
DLB$^{1}$		& 0.293$\pm$0.009	&	0.747$\pm$0.008 		&	1--192		& 256		\\
OCTI$^{1}$		& 0.273$\pm$0.004	&	0.752$\pm$0.003 		&	14--208		& 829		\\
OCTII$^{1}$		& 0.254$\pm$0.004	&	0.764$\pm$0.003			&	16--203		& 829		\\
SYD$^{1}$		& 0.268$\pm$0.004	&	0.758$\pm$0.003			&	8-220		& 799		\\
CoRoT$^{2}$ 	& 0.28$\pm$0.02		&   0.75$\pm$0.01			&	2--100		& 1827		\\
\hline
MS+RG$^{3}$		& 0.263$\pm$0.009	&	0.772$\pm$0.005			&	15--4500	& 55		\\
\hline
\end{tabular} 
\end{center}
$^{\mathrm{1}}$\,\kep\ red giants: A2Z - \citet{mathur10}, CAN - \citet{kall_comp}, 
COR - \citet{mosser09}, DLB - Buzasi et al. (unpublished), OCTI \& OCTII - \citet{hekker09b}, 
SYD - \citet{HSB09}. \newline
$^{\mathrm{2}}$\,\textit{CoRoT} red giants, see \citet{mosser10}. \newline
$^{\mathrm{3}}$\,Main-sequence (MS) and red giant (RG) stars, see \citet{stello09} and 
references therein.
\label{tab:dnunumax} 
\end{table}

\begin{figure}
\begin{center}
\resizebox{\hsize}{!}{\includegraphics{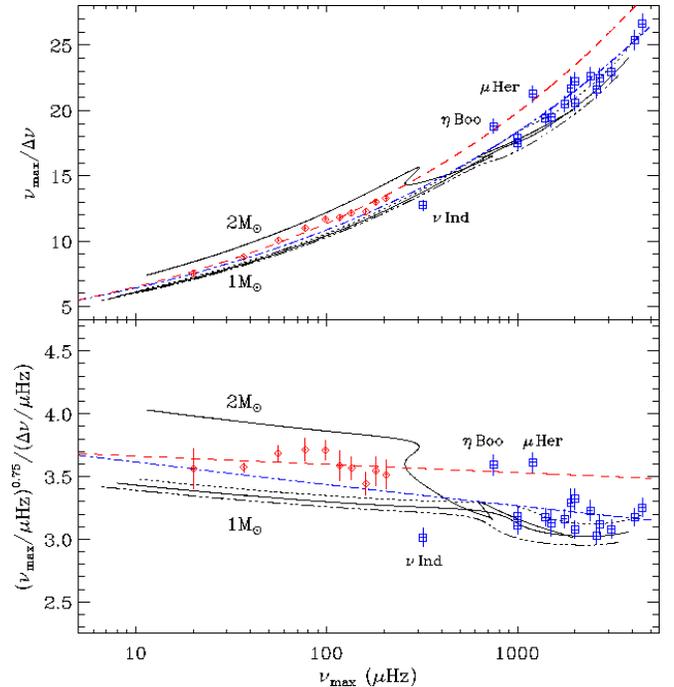}}
\caption{\textit{Upper panel:} Solar-metallicity (Y=0.283, Z=0.017) evolutionary tracks (solid lines) together 
with the relation by \citet{stello09} (blue dashed-dotted line) and the relation fitted to the 
\kep\ data (red dashed line). Blue squares show the main-sequence and sub-giant 
stars used by \citet{stello09}, and red diamonds the \kep\ observations as shown in 
Figure \ref{fig:02} in bins of 20\,\muHz. The black dashed triple-dotted and dotted lines show the solar-mass 
tracks for chemical compositions of (Y,Z) = (0.291,0.009) and (Y,Z) = (0.265,0.035), respectively. 
\textit{Lower panel:} Same as the upper panel, but with the luminosity dependence of the 
ordinate removed by raising \numax\ to the power of 0.75 (see Equation (\ref{equ:dnunumax2})).}
\label{fig:03}
\end{center}
\end{figure}

As noted by \citet{mosser10}, the difference between the relation for red giants and the one 
including main-sequence stars by \citet{stello09} is small but significant. To illustrate 
this difference we plot the relations 
together with observations and evolutionary tracks in Figure \ref{fig:03}. Note that \numax\ and \Dnu\ 
for the models have been calculated using equations (\ref{equ:nmax}) and (\ref{equ:dnu}). Compared to 
Figure \ref{fig:02}, we replaced the ordinate in the upper panel by the ratio \numax/\Dnu, which 
scales with the 
radial order of maximum power. The lower panel shows this ratio with \numax\ raised to the 
power of 0.75, which better illustrates the mass dispersion since in this plot, constant masses correspond to almost 
horizontal lines (see Equation (\ref{equ:dnunumax2})).
The \kep\ stars are shown in bins of 20\,\muHz\ (red diamonds).
Note that the red giants used by \citet{stello09} have been omitted for clarity, but do not 
differ significantly from the \kep\ sample. The uncertainties for the main-sequence sample have 
been collected from the literature where available or were set to typical values of 3\% for \numax\ and 
1\% for \Dnu.

The scatter about the \Dnu-\numax\ relation is considerably larger than the 
measurement uncertainties (which are $\sim$1\% for \Dnu\ in our sample) and for red giants is
mainly caused by the spread of stellar masses \citep{kallinger10}. While Figure \ref{fig:03} 
shows that this spread is significant on the red-giant branch, it can be 
seen that the evolutionary tracks (solid lines) almost overlap for main-sequence stars. 
This effect causes a different fit to the relation between \Dnu\ and \numax\ for a sample 
consisting only of red giants compared to when main-sequence stars 
are included, which is 
clearly reflected by the observations and the determined fits shown in Figure \ref{fig:03}. 

The dotted and dashed triple-dotted lines in Figure \ref{fig:03} illustrate the effect of changing the 
metallicity for the 1\,$M_{\sun}$ track. The effect is
relatively small on the giant branch \citep{kallinger10}, but more significant for less evolved stars. 
In fact, both main-sequence stars significantly above the \citet{stello09} relation, $\eta$\,Boo and 
$\mu$\,Her, are observed to be metal-rich \citep{carrier_etaboo,yang_muher}, while the sub-giant 
$\nu$\,Indi which falls below the relation is metal-poor \citep{bedding_nuindi}. Hence, these 
observations are qualitatively in agreement with expectations from evolutionary models 
combined with scaling relations.

A detailed investigation of these effects using pulsation models (which would be necessary to 
quantify mass and metallicity changes in Figure \ref{fig:03} in absolute terms) is beyond the scope of 
this paper. However, our results show that the observed difference 
in the \Dnu-\numax\ relation for red giants compared to when main-sequence stars are included is 
consistent with scaling relations. A more detailed investigation of the relation for main-sequence 
stars using \kep\ short-cadence data will be presented in a forthcoming paper.

\vspace{0.15cm}
\subsection{Population effects in the \numax\ and \Dnu\ distributions}
The large number and diversity of red giants for which solar-like oscillations have been detected 
enables us to identify stellar populations using global oscillation parameters. Due to 
the straightforward relationship between oscillation parameters and fundamental parameters (see 
equations (\ref{equ:nmax}) and (\ref{equ:dnu})), comparisons with models could allow us to draw 
some conclusions about the star-formation history in the \kep\ field. Such studies have 
already been performed for red giants observed by \textit{CoRoT} 
\citep{miglio09,yangmeng}.

\begin{figure*}
\begin{center}
\resizebox{8cm}{!}{\includegraphics{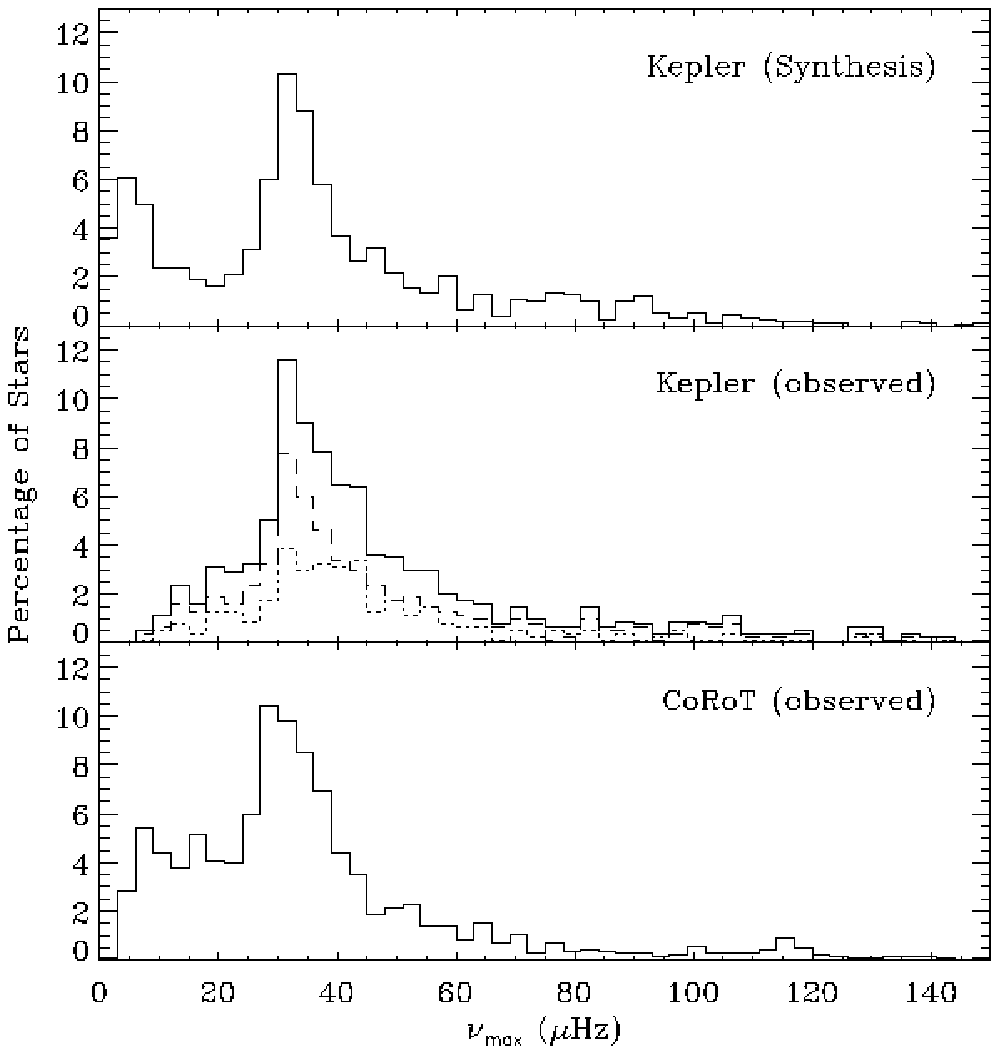}}
\hspace{0.5cm}
\resizebox{8cm}{!}{\includegraphics{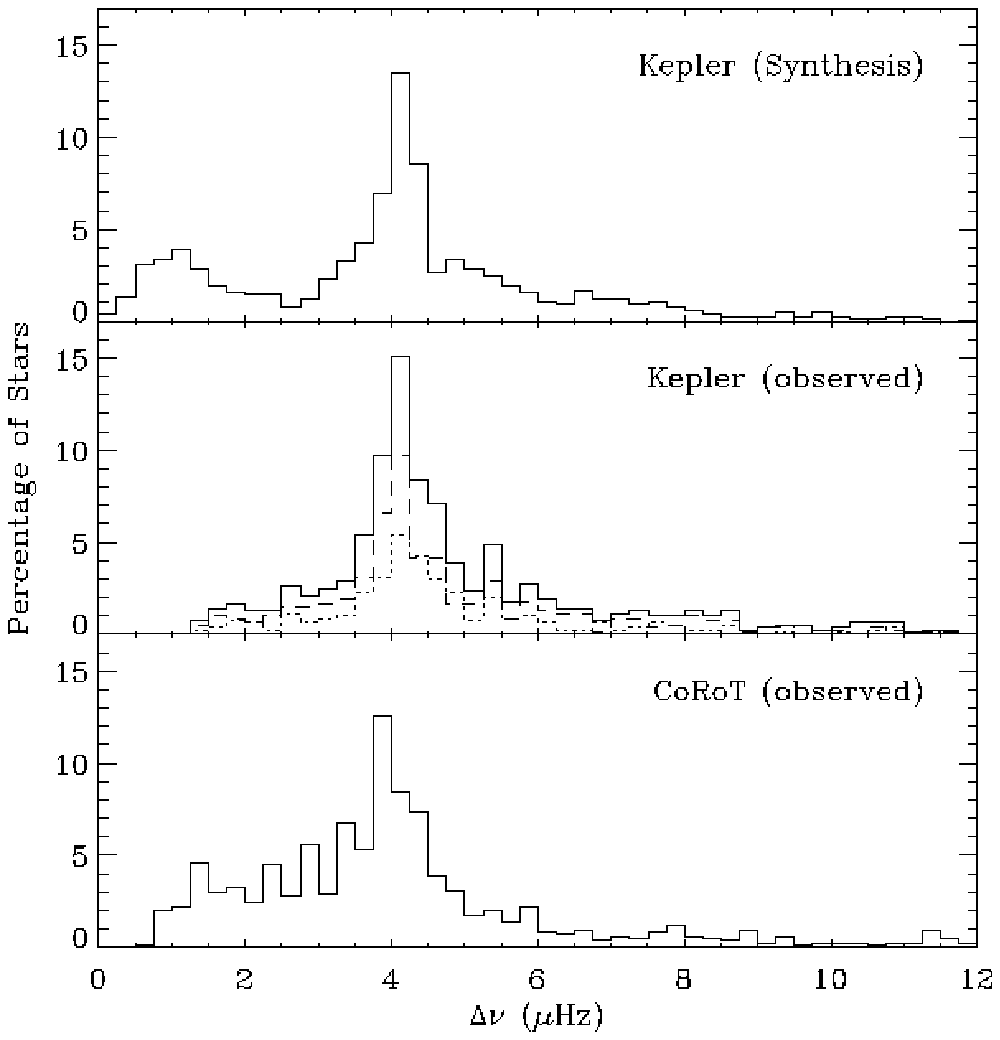}}
\caption{Histograms of \numax\ (left panels) and \Dnu\ (right panels) in 
percent comparing a synthetic stellar population for the \kep\ FOV (top panels), the observed 
distributions in the \kep\ sample (middle panels) and the observed distributions in the 
\textit{CoRoT} sample (bottom panels). 
The dashed and dotted lines in the middle panel separately show the 
distributions of the astrometric and asteroseismic sample, respectively. Note that the 
comparison for the \kep\ sample should be restricted to stars with $\numax\gtrsim 10\,\muHz$ 
($\Dnu\gtrsim 1.5\,\muHz$).}
\label{fig:histos}
\end{center}
\end{figure*}

Before such a comparison is made, it is important to consider potential biases in the sample 
of \kep\ red giants. Firstly, limitations on the bandwidth of spacecraft communications restrict 
the number of stars for which data can be obtained. Hence, absolute star 
counts for a given field and magnitude range will be incomplete. Secondly, about two thirds of 
the \kep\ red-giant sample have been chosen as 
astrometric reference stars \citep{batalha,monet}. These have been selected to be distant, bright but
unsaturated giants ($T_{\rm eff,KIC}<5400$\,K, $\log g_{\rm KIC}<3.8$, $R/R_{\sun,\rm KIC}>2$) 
in the \kep\ magnitude range 11.0--12.5. Through the remainder of this paper we will 
refer to these stars as the astrometric sample, and to the remaining stars as the asteroseismic 
sample. Thirdly, as was the case for the 
\textit{CoRoT} sample, the \kep\ sample excludes stars pulsating at frequencies that are 
too low to be resolved with the given time base available at this stage of the mission ($\numax\lesssim 10\,\muHz$).

To obtain a first qualitative analysis of the observed distributions of the \kep\ sample, we 
calculated a synthetic stellar 
population the same way as \citet{miglio09}, using the stellar synthesis code {\sc{TRILEGAL}} 
\citep{trilegal}. The input parameters were identical to the simulations performed by \citet{miglio09} for 
the \textit{CoRoT} sample, with the exception of restricting the simulation to 
a 10 deg$^{2}$ field centered on the Kepler FOV ($\alpha$=290.7$^{\circ}$, $\delta$=44.5$^{\circ}$) 
in the magnitude range 8 -- 13 mag in the \kep\ bandpass. A constant star-formation rate was assumed.

Figure \ref{fig:histos} compares the distributions of the synthetic \kep\ population with the 
observed one, together with the \textit{CoRoT} sample \citep{mosser10}. Note that the latter 
includes stars from both fields of view of \textit{CoRoT}, one directed to the galactic 
center, the other to the galactic anti-center.  As for the \textit{CoRoT} sample, we observe a maximum 
at \numax$\sim$\,30\,\muHz, corresponding to the red clump. These are low-mass ($< 2 M_{\sun}$) stars, 
for which He-core burning occurs at similar luminosities (and hence similar \numax), forming a dominant 
population on the giant branch. \citet{miglio09} noted that for the \textit{CoRoT} stars, the maximum appears 
at a lower value of \numax\ (higher luminosity) than for the model. This is not the case 
for the \kep\ field.

The dashed and dotted lines in the middle panel show 
the distributions of the astrometric and asteroseismic sample separately, to illustrate 
potential biases introduced in the overall distribution by combining these distinctly selected sets 
of stars. We see that the red clump maximum is predominantly formed by the astrometric sample, 
while the asteroseismic sample contributes more to slightly higher \numax\ values. 
This suggests that the astrometric sample is largely unbiased, while the 
broadening of the red clump peak compared to the model is potentially caused by 
selection bias in the asteroseismic sample.

Apart from the red clump, a second and much broader component in the synthetic population can be 
identified in the interval $\numax\simeq 40-110\,\muHz$ and $\Dnu\simeq 5-10\,\muHz$. 
This consists of more massive stars 
that, compared to the red clump, occupy lower luminosities (and hence higher \numax) over a 
wider range when they settle as 
He-core burning stars. This component is referred to as the secondary clump by \citet{girardi_secclump}. 
A comparison with the histogram of the \kep\ sample shows that the 
observations qualitatively reproduce the distributions for both higher \numax\ and \Dnu\ values.

\begin{figure}
\begin{center}
\resizebox{\hsize}{!}{\includegraphics{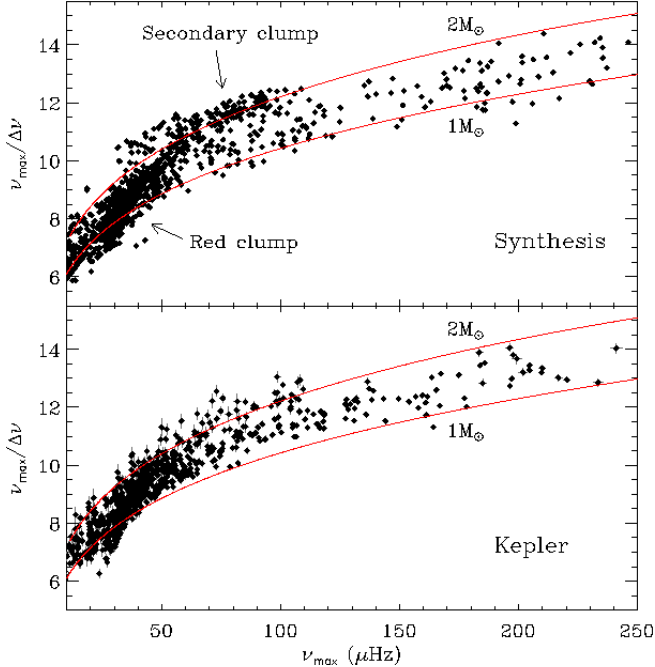}}
\caption{Ratio of \numax/\Dnu\ as a function of \numax\ for the synthetic population (upper panel, 
1530 stars) and the \kep\ red-giant sample (lower panel, 801 stars). Red lines 
show solar-metallicity (Y=0.283, Z=0.017) 
{\sc{ASTEC}} evolutionary tracks with masses as indicated in the plots. The red clump and 
secondary clump populations are indicated for the synthetic population.}
\label{fig:dists}
\end{center}
\end{figure}

\begin{figure}
\begin{center}
\resizebox{\hsize}{!}{\includegraphics{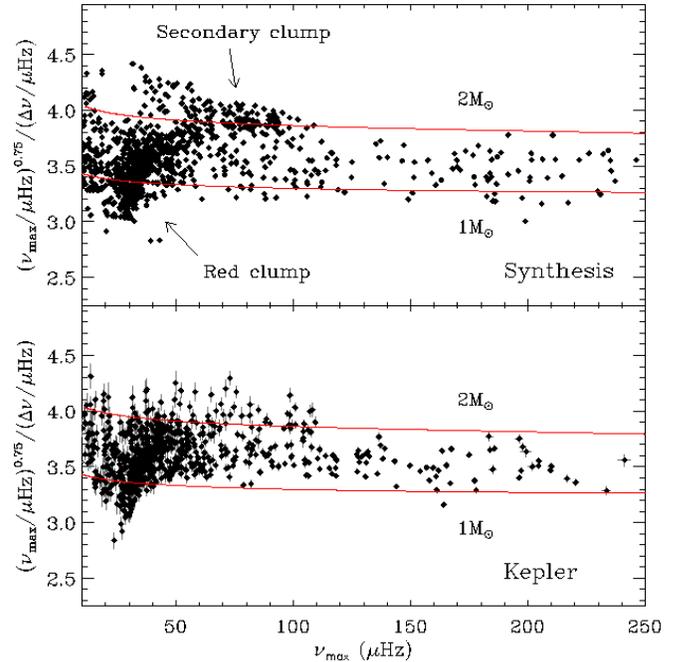}}
\caption{Same as Figure \ref{fig:dists}, but with the luminosity dependence of the 
ordinate removed by raising \numax\ to the power of 0.75 (see Equation (\ref{equ:dnunumax2})).}
\label{fig:dists2}
\end{center}
\end{figure}

To investigate this further, Figure \ref{fig:dists} shows the 
ratio \numax/\Dnu\ as a function of \numax\ for the synthetic population and for the \kep\ 
observations. The solid lines are evolutionary tracks of different masses. 
Figure \ref{fig:dists2} shows the same plot but with the luminosity dependence 
on the ordinate removed by raising \numax\ to the power of 0.75.
In both figures, the separation of the red clump and the secondary clump is obvious in 
the population synthesis (upper panels). For the 
observations (lower panels) we can identify a group of stars with  $M>2\,M_{\sun}$ extending up to 
$\numax\sim110\,\muHz$, as expected for the secondary clump population. 

The qualitative agreement between the distributions is encouraging 
and the excess of stars with $\numax=40-110\,\muHz$\ and $M>2\,M_{\sun}$ suggests that 
the \kep\ sample includes stars belonging to the secondary clump population. As pointed out by 
\citet{girardi_secclump}, the detection of these stars should allow us to probe 
important physics, such as convective-core overshooting, and help to put tight 
constraints on the recent star-formation history in the galaxy. The detailed modelling required 
for these inferences is beyond the scope of this paper.

\subsection{Small frequency separations}

As discussed in Section \ref{sec:spacings}, the small frequency separations of main-sequence 
stars depend on the sound-speed gradient in the stellar core and are  
sensitive to the evolutionary state of a star. The excellent quality of the \kep\ data 
allows us to measure the small separations in an unprecedented number of stars covering a wide 
range of evolutionary states on the giant branch.

To measure the small separations, we first examined the \'{e}chelle diagram of 
each star using the power spectrum in the frequency range $\numax\pm 5\Dnu$ and then manually fine-tuned
the large separation to make the $l=0$ ridge vertical. 
The adjustment to \Dnu\ was typically a few tenths of a microhertz. Stars for which no unambiguous 
identification of $l=0$ and 2 could be found were discarded from this analysis. The adjusted \'{e}chelle 
diagram was then collapsed, and a Gaussian function was fitted to each mode ridge 
that was identified. The center of the fitted Gaussian was taken as the 
position of that ridge. Using this technique we were able to measure the $l=0$ and 2 ridges 
for 470 stars. Of these, we measured the $l=1$ ridge for 400 stars and the $l=3$ ridge 
for 45 stars. The uncertainties were determined from extensive simulations of artificial \kep\ data 
that were analyzed using the same method. Before discussing the separations 
of the ridges, we first consider their absolute positions.

\subsubsection{Variation of $\epsilon$}

\begin{figure}
\begin{center}
\resizebox{\hsize}{!}{\includegraphics{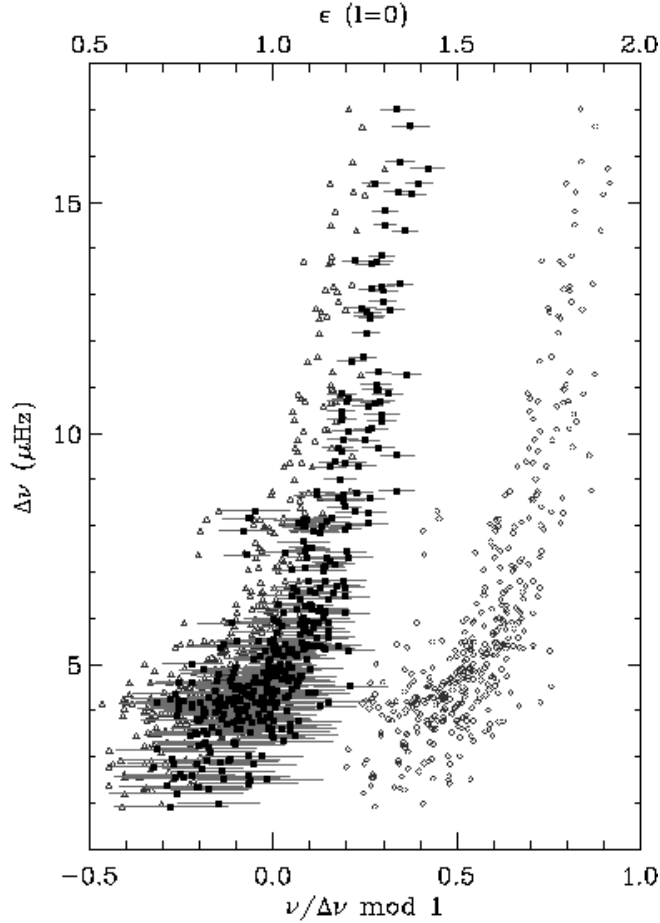}}
\caption{Mode ridge centroids as measured from the folded and collapsed power spectrum for $l=0$ (black squares), 
$l=1$ (grey diamonds) and $l=2$ (grey triangles). Note that the upper abscissa label shows 
$\epsilon$ for $l=0$ modes. Error bars are only shown for $l=0$ for clarity.}
\label{fig:epsilon}
\end{center}
\end{figure}

The measured mode ridge centroids allow us to investigate the parameter
$\epsilon$ in Equation (\ref{equ:asymt}). Guided by the Sun, for which $\epsilon\simeq1.5$ 
\citep[see, e.g., Equation (11) in][]{kb95}, we have plotted the centroid positions in 
Figure \ref{fig:epsilon}, which shows that $\epsilon$ is a function of \Dnu.

Figure \ref{fig:epsilon} also shows a considerable spread in $\epsilon$ for a given \Dnu, 
in particular for stars with $\Dnu<5\,\muHz$.
While the spread is of the same order as the estimated uncertainties, we have tested whether some 
of it might be related to physical 
properties of the stars. We have investigated this by correlating 
$\epsilon$ with \numax/\Dnu\ which, as shown in Figure \ref{fig:dists}, is sensitive to stellar 
mass. No clear correlation could be found.

The relation between $\epsilon$ and \Dnu\ (and hence also \numax) implies that $\epsilon$ is a 
function of fundamental parameters. If this relation can be quantified 
for less-evolved stars, clear mode ridge identifications as presented in this paper could potentially be 
used to predict $\epsilon$ for other stars. As suggested by \citet{bedding_scaling}, such comparisons 
can be of considerable help in cases where the mode ridge identification is difficult 
due to short mode lifetimes or rotational splitting 
\citep[see, e.g.,][]{app08,benomar09,garcia_hd181906,kallinger_hd49933,bedding_procyon}.

Meanwhile, we can already use the ensemble results in Figure \ref{fig:epsilon} to suggest
ridge identifications for the four \textit{CoRoT} red giants discussed by \citet{hekker10b}. 
These stars have \Dnu\ in the range 3.1 to 5.3\,\muHz\ and so we expect $\epsilon$ to be 1.0 or slightly 
less.  Looking at the \'{e}chelle diagrams of these stars \citep[see Figures  8, 10, 12 and 14 in][]{hekker10b}, 
we indeed see in all cases that one of the two ridges falls there.
In each of these diagrams, this indicates it is the left-hand ridge that
corresponds to $l=1$ and the other to $l=0$.

\subsubsection{C-D diagrams}
\begin{figure}
\begin{center}
\resizebox{\hsize}{!}{\includegraphics{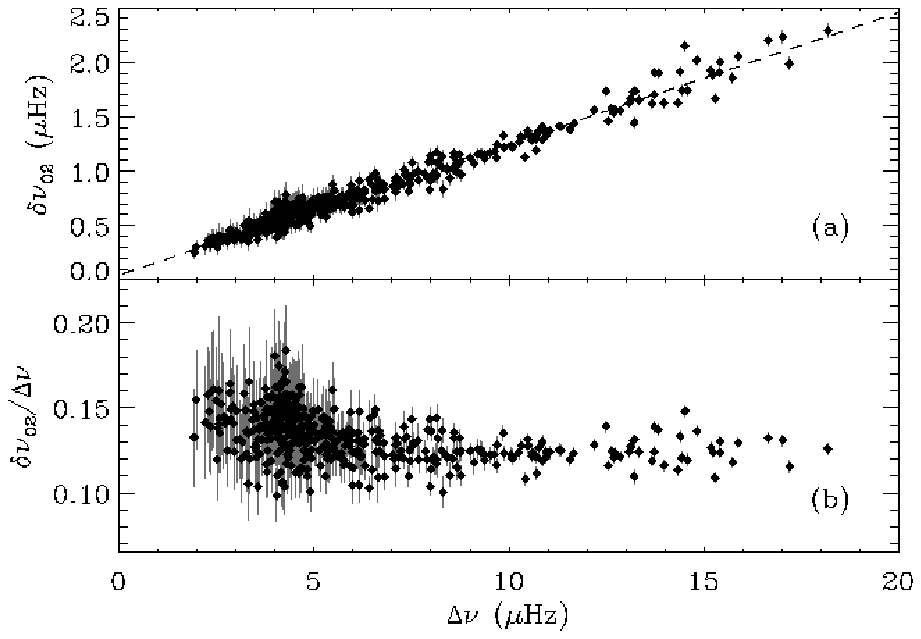}}
\resizebox{\hsize}{!}{\includegraphics{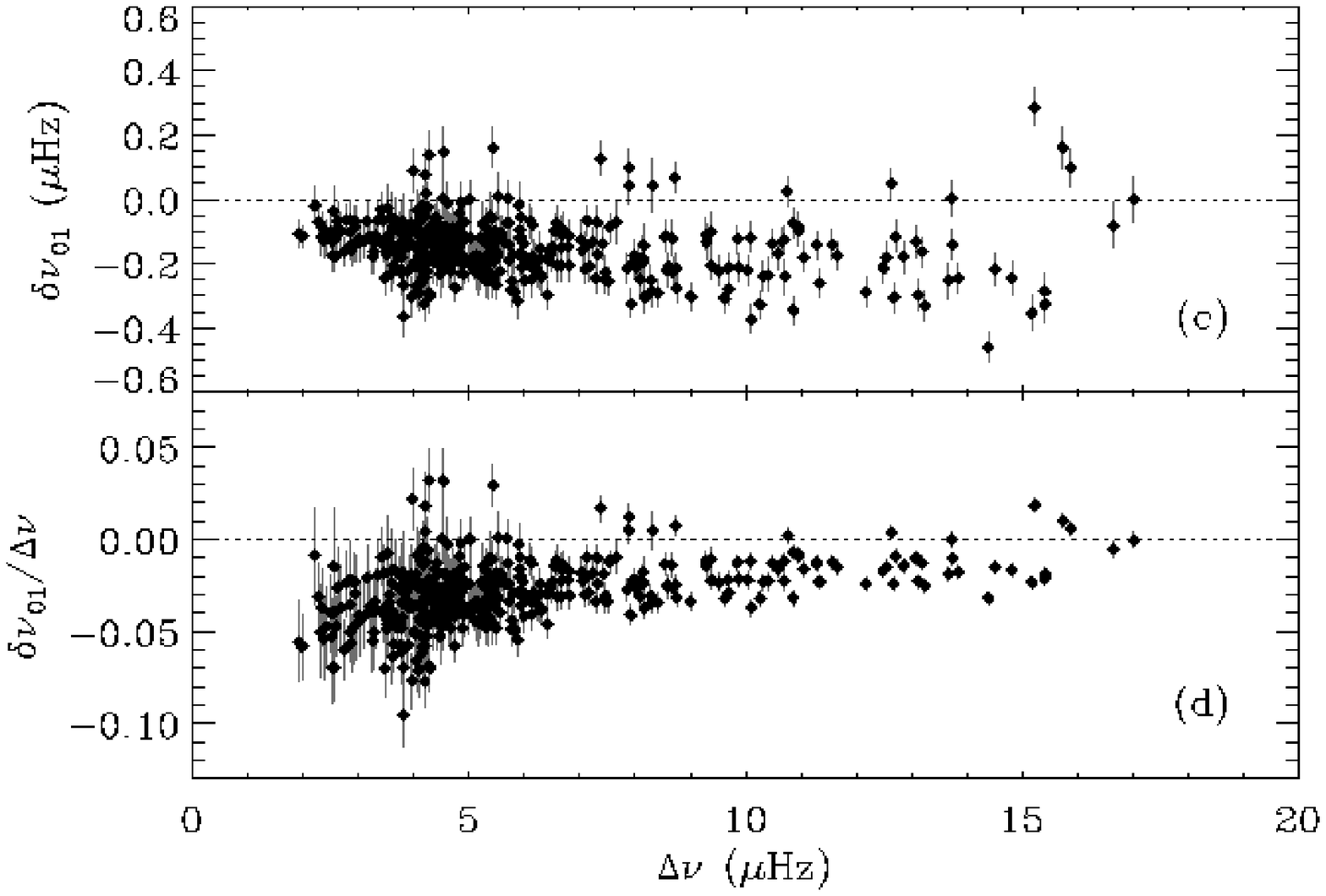}}
\resizebox{\hsize}{!}{\includegraphics{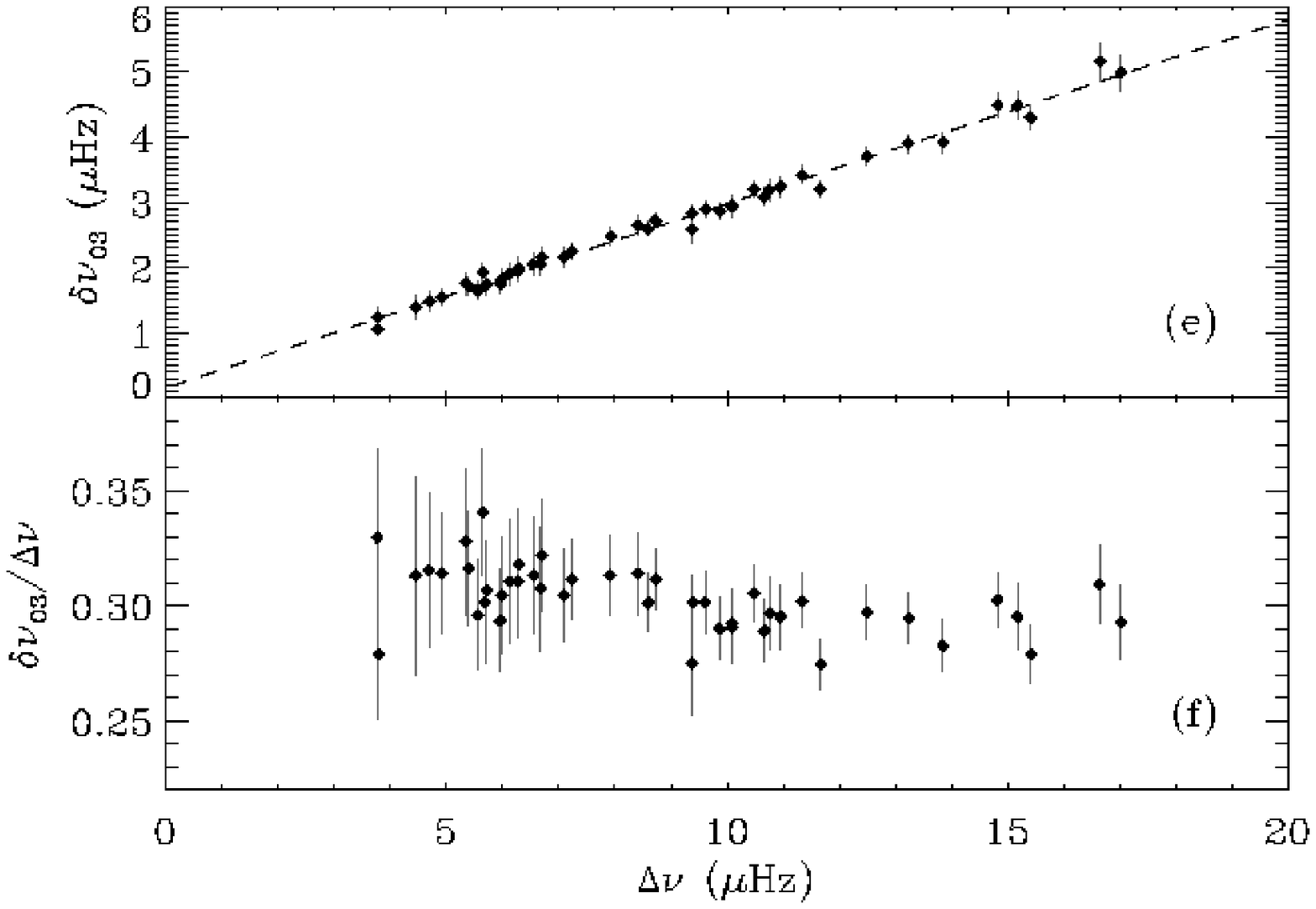}}
\caption{C-D diagram of \dnu{0\rm l}\ versus \Dnu\ for $l=2 $ $(a)$, $l=1$ $(c)$ and $l=3$ $(e)$ 
and frequency separation ratios of \dnu{0\rm l}/\Dnu\ for $l=2$ $(b)$, $l=1$ $(d)$ and $l=3$ $(f)$. Dashed lines in 
panels $(a)$ and $(e)$ show linear fits to the data.}
\label{fig:smallsp}
\end{center}
\end{figure}

Using the measured mode ridge centroids, we were able to derive the small separations \dnu{02}, 
\dnu{01}\ and \dnu{03}\ for 470, 400 and 45 stars, respectively. The results are shown 
in so-called C-D diagrams \citep{ChD88b} in Figure \ref{fig:smallsp}. We also plot the frequency 
separation ratios \dnu{0l}/\Dnu\ which, according to models, are expected to be largely insensitive to surface 
layer effects \citep{roxburgh,floranes,mazumdar,chaplin05}.

We can see in Figure 
\ref{fig:smallsp}a that \dnu{02}\ is an almost fixed fraction of \Dnu, which confirms the findings by 
\citet{BHS10} for low-luminosity red giants. We obtain the following relation for the full range of \numax:

\begin{equation}
\delta\nu_{02} = (0.121 \pm 0.001) \Delta\nu + (0.047 \pm 0.008) \: ,
\end{equation}

\noindent
in agreement with the relation found by \citet{BHS10}. We see a deviation from this linear 
relationship for stars with $\Dnu\lesssim10\,\muHz$ 
($\numax\lesssim120\,\muHz$), which can be seen in Figure \ref{fig:smallsp}b as an increase of the 
ratio $\dnu{02}/\Dnu$. Additionally, we observe an increased spread of \dnu{02}, particularly 
for stars with $\Dnu\sim4\,\muHz$. While most of this spread in the red clump is due to the 
much larger number of stars (see Figure \ref{fig:histos}) and 
the larger uncertainty in determining \dnu{02}, we tested whether it could also 
be related to physical properties of the stars. C-D diagrams for red giants calculated 
from stellar models (T. R. White et al., in preparation) show that the 
expected range in $\dnu{02}/\Dnu$ for $M=1-2\,M_{\sun}$ and solar metallicity in non He-core 
burning models is about 0.02, which is roughly comparable to the range of values observed in the data 
outside the red clump.
Indeed, a comparison of stars with different $\dnu{02}$ in the lower panel of Figure \ref{fig:dists} 
has shown the spread can be partially explained by a spread in stellar masses, with lower-mass 
stars showing higher \dnu{02}\ values. More quantitative conclusions about the mass spread 
will be possible once the measurement uncertainties are reduced by the collection of more data.

Figure \ref{fig:smallsp}c shows that \dnu{01}\ is negative for almost all red giants, 
confirming the findings by \citet{BHS10}. As for $l=2$, we 
observe a trend (but with opposite sign) in the frequency separation ratio \dnu{01}/\Dnu\ 
with \Dnu, which can be seen in Figure \ref{fig:smallsp}d. 
The decrease of \dnu{01}/\Dnu\ and the increase of \dnu{02}/\Dnu\ appear to affect stars in the same 
range of \Dnu\ and by a similar amount ($\sim$5\%). Using stars common to both 
samples, we calculated a correlation coefficient of $-0.1$. While this indicates that the observed 
trends are statistically almost uncorrelated, we cannot at this point exclude that this is due 
to the large uncertainty of measuring \dnu{01}.

As shown by \citet{BHS10}, the \kep\ data allow the detection of $l=3$ modes in red giants.
We were able to measure the small separation \dnu{03} for 45 stars and a linear fit to the relation 
with \Dnu, shown in Figure \ref{fig:smallsp}e, yields:

\begin{equation}
\delta\nu_{03} = (0.282 \pm 0.005) \Delta\nu + (0.16 \pm 0.04) \: .
\end{equation}

\noindent
Figure \ref{fig:smallsp}f shows the ratio \dnu{03}/\Dnu, 
which shows some evidence for an increase of \dnu{03}\ with decreasing \Dnu. While the error bars appear to be too 
large to make any firm statements on this variation, the observed scatter
suggests that the uncertainties are overestimated. Given that the relative amplitude of $l=3$ modes 
is not well known, it seems likely that the strength of $l=3$ modes has been considerably 
underestimated in the simulations used to derive the uncertainties.

\subsubsection{Ensemble collapsed \'{e}chelle diagram}

\begin{figure}
\begin{center}
\resizebox{\hsize}{!}{\includegraphics{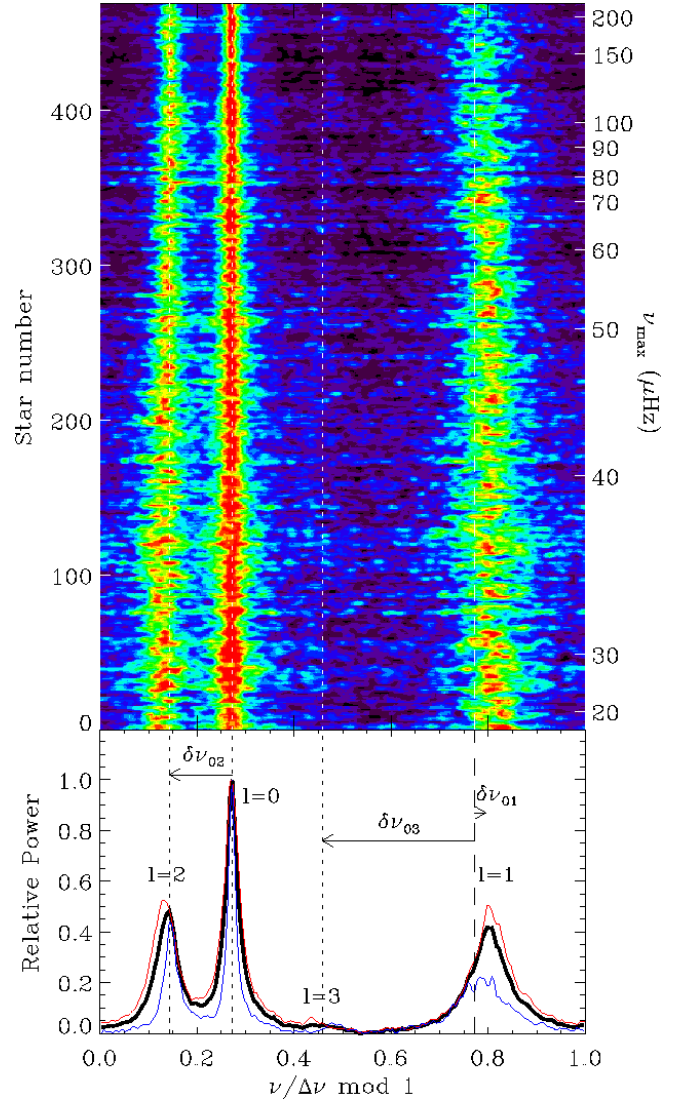}}
\caption{\textit{Upper panel:} Ensemble collapsed \'{e}chelle diagram of 
470 red giants. Note that the contour threshold has been lowered to 40\% of the maximum value 
and that each row shows the folded and collapsed power spectrum of one star. The right ordinate 
marks \numax\ values for selected rows.
\textit{Lower panel:} Upper panel collapsed along the entire range of \numax\ (thick black line) 
as well as in subsets $\numax>100\,\muHz$ (thin blue line) and $\numax<50\,\muHz$ (thin red line). 
Ridge identification 
and definitions of small separations used in this paper are indicated. In both panels, the dotted 
lines mark the centers of the $l=0$, 2 and 3 ridge and the dashed line shows the midpoint of 
adjacent $l=0$ modes.}
\label{fig:foldedps}
\end{center}
\end{figure}

In order to display the separations of the mode ridges for all stars, we use the measured values of
$\epsilon$ to shift and align the folded and collapsed power spectra in an ensemble collapsed 
\'{e}chelle diagram. The result is shown in the upper
panel of Figure \ref{fig:foldedps}. Note that this diagram is slightly different from the scaled \'{e}chelle 
diagram presented by \citet{BHS10} in that we have \textit{shifted} the folded 
and collapsed power spectra rather than \textit{scaling} frequencies. The thick solid line in the
lower panel shows the upper panel summed along the full range of \numax. It clearly reveals the 
presence of ridges with $l=0,1,2$, and also $l=3$. 
The power for $l=3$ is weaker than found by
\citet{BHS10}, who considered a smaller sample of low-luminosity giants with high signal-to-noise.  
We attribute this to the larger 
(and less biased) sample of stars in this paper, including stars with undetectable $l=3$ modes, 
as well as to the fact that the $l=3$ ridge has a slight tilt, making its power spread out in the
lower panel of Figure \ref{fig:foldedps}.

The trends of the frequency separation ratios observed in Figure \ref{fig:smallsp} are clearly 
visible in Figure \ref{fig:foldedps}. As \numax\ (and hence \Dnu) decreases, the $l=2$ 
ridge slopes to the left while the $l=1$ ridge slopes to the right, corresponding to an increase 
of \dnu{02}/\Dnu\ and a decrease of \dnu{01}/\Dnu. 
These features are also clearly shown by the thin red and blue lines in the lower panel of 
Figure \ref{fig:foldedps}, which correspond to subsets 
with $\numax<50\,\muHz$ and $\numax>100\,\muHz$, respectively.
The red and blue lines also show evidence for a slope of the $l=3$ ridge, corresponding to an 
increase of \dnu{03}\ as observed in Figure \ref{fig:smallsp}f.

We note that the \textit{relative} width of the $l=0$ ridge shown in Figure \ref{fig:foldedps} 
(measured in terms of \Dnu) decreases with increasing \numax. In fact, the \textit{absolute} width 
of the ridge (in \muHz), which is 
determined by the frequency resolution, mode lifetime and curvature, remains roughly constant over 
the range of stars considered. As found by \citet{BHS10} for low-luminosity red giants, the $l=1$ ridge 
is significantly broader than the others. This was interpreted as a direct confirmation of 
theoretical results by
\citet{dupret}, which predicted complicated power spectra due to less efficient trapping of mixed modes 
in the cores of low-luminosity red giants. The models by \citet{dupret} also 
predicted that this effect should become less pronounced for higher luminosity red giants. 
Indeed, we observe that the relative width of the $l=1$ ridge shown in Figure \ref{fig:foldedps} 
remains roughly constant, indicating that its absolute width is significantly lower 
for high-luminosity stars than for low-luminosity stars.

\begin{figure}
\begin{center}
\resizebox{\hsize}{!}{\includegraphics{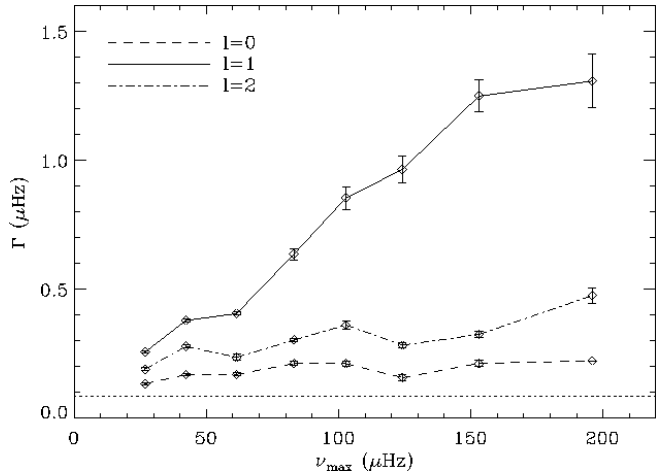}}
\caption{Absolute width of the $l=0$ (dashed line), $l=1$ (solid line) and 
$l=2$ (dashed-dotted line)
ridge as a function of \numax. The dotted line marks the formal frequency resolution of the data 
(0.08\,\muHz).}
\label{fig:widths}
\end{center}
\end{figure}

To quantify this, we measured the absolute widths of the $l=0$, 1 and 2 ridges in several bins of 
\numax, and show the result in Figure \ref{fig:widths}. Note that we have accounted for 
possible artificial broadening of the ridges due to Nyquist effects by excluding the 
two stars in our sample with $\numax>230\muHz$ from this calculation. 
As can be seen in Figure \ref{fig:widths}, the widths of the $l=0$ and 2 ridges remain 
roughly constant over the range of \numax, while the width of the $l=1$ ridge increases significantly. 
Our observation of a much narrower $l=1$ ridge for high-luminosity 
stars therefore provides further confirmation of the results by \citet{dupret}. 

\subsection{Pulsation amplitudes}

Amplitudes of solar-like oscillations provide valuable information to test models of convection 
(see Section \ref{sec:amplitudes}). Equations (\ref{equ:nmax}), (\ref{equ:amps}) and (\ref{equ:amps2}), 
in combination with effective temperatures, can be used to test theoretical scaling relations for 
amplitudes \citep[see, e.g.,][]{stello06}, as follows:

\begin{equation}
A_{\lambda} \propto T_{\rm eff}^{3.5s-r} \frac{\nu_{\rm max}^{-s}}{\lambda} \: .
\label{equ:amprel}
\end{equation}

\begin{figure}
\begin{center}
\resizebox{\hsize}{!}{\includegraphics{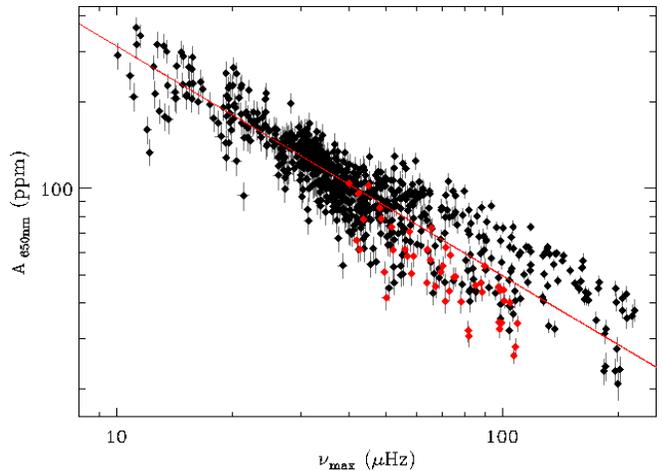}}
\caption{Pulsation amplitude versus \numax\ observed in the \kep\ bandpass 
($\lambda=650$\,nm). The red solid line shows a power law using a slope of 
$-0.8$. Red symbols show all stars with $\numax>40\,\muHz$ and $M>2M_{\sun}$ identified in 
Figures \ref{fig:dists} and \ref{fig:dists2}.}
\label{fig:amps}
\end{center}
\end{figure}

\noindent
As shown by \citet{mosser10} for the \textit{CoRoT} red giants, 
the relation between \numax\ and the pulsation amplitudes indeed follows a power law and this
is shown for our \kep\ sample in Figure \ref{fig:amps}. We have investigated the slope of the power law 
using amplitudes from four different methods (A2Z, COR, OCT and SYD, see Table 1 for 
references) and find on average:

\begin{equation}
A_{\rm 650nm} \propto \nu_{\rm max}^{-0.8} \: .
\label{equ:ampobs}
\end{equation}

\noindent
As can be seen from Figure \ref{fig:amps}, the measured amplitudes show considerable structure, 
in particular a lack of low amplitudes for $\numax\gtrsim110\,\muHz$. This feature is found 
consistently in all methods and therefore appears to be intrinsic. Some low-amplitude 
stars do appear at $\numax\gtrsim180\,\muHz$, but we note 
that these amplitude estimates are rather uncertain due to the difficulty of estimating the 
background noise for stars oscillating near the Nyquist frequency.
To investigate this lack of low amplitudes, we plot all stars that we have 
tentatively identified as secondary clump stars (using Figures \ref{fig:dists} and \ref{fig:dists2}, 
setting $\numax>40\,\muHz$ and $M>2M_{\sun}$) in Figure \ref{fig:amps} as red symbols. 
We observe that these stars have systematically low amplitudes. We conclude that, for a given 
\numax\ in the range between $40-110\,\muHz$, stars with lower \Dnu\ (and therefore higher 
mass, see Equation (10)) oscillate with lower amplitudes than other stars with the same \numax. 
Assuming that this also holds for low-luminosity red giants, the lack of low amplitude 
stars for $\numax\gtrsim110\,\muHz$ could then possibly be explained by the 
relatively fast evolution of more massive stars near the bottom of the red-giant branch compared 
to $\sim 1\,M_{\sun}$ stars.

Both the observation of low amplitudes for stars with lower \Dnu\ in 
the range $\numax\simeq 40-110\,\muHz$ and the lack of low amplitude stars for $\numax\gtrsim110\,\muHz$ 
indicate a significant dependence of the amplitude-\numax\ relation on the stellar mass. Since such a 
dependence is not seen in Equation (\ref{equ:amprel}), this
suggests a revision of the scaling relation in Equation (\ref{equ:amps}).

The overall distribution of the measured amplitudes 
in Figure \ref{fig:amps} shows considerable additional scatter, which is possibly related to physical 
effects such as metallicity \citep{samadi10a,samadi10b} in 
combination with measurement uncertainties. Tests have shown that the measurement error for 
amplitudes is dominated by the specific method and model used to subtract the background signal due 
to granulation and stellar activity. Further work is needed to identify possible systematic effects 
for the entire range of \numax, which will then allow us to improve the relation between 
the oscillation amplitude and \numax\ using a large sample of stars on the giant branch.

\section{Summary and conclusions}

Using data from the first four months of the \kep\ mission we have studied various aspects of 
solar-like oscillations in $\sim$\,800 red giants. Our conclusions can be summarized as follows:

\begin{enumerate}

\item The \Dnu-\numax\ relation using red giants is significantly different from the relation
when main-sequence stars are included. This difference can be explained by evolutionary models and 
scaling relations, which show that the \Dnu-\numax\ relation is mainly sensitive to stellar masses for red giants and 
to metallicity for main-sequence stars. 

\item The \numax\ and \Dnu\ distributions are in qualitative agreement with a simple model of the 
stellar population.  We observed that \numax\ for the red clump stars in the Kepler field 
is slightly higher than for the \textit{CoRoT} fields, possibly due to the different stellar 
populations that are observed.
We identified several 
stars with $\numax=40-110\,\muHz$ and $M>2M_{\sun}$, in agreement with a secondary clump population. 

\item The quantity $\epsilon$ in the asymptotic relation is a function of \Dnu.
If the relation between $\epsilon$ and fundamental parameters can 
be quantified and is also valid for main-sequence stars, it could be used to study surface layer 
effects or for mode identification in stars where ridges cannot be clearly identified
due to short mode lifetimes or rotational splitting. 
We demonstrated the potential of the observed relation by providing mode identifications for four 
red giants observed by \textit{CoRoT}.

\item We presented C-D diagrams for $l=1$, 2 and 3 and show that frequency separation ratios of 
\dnu{02}/\Dnu\ and \dnu{01}/\Dnu\ reveal opposite trends as a 
function of \Dnu. We observed a spread in the small separations, in particular for stars
in the red clump, and find evidence that this is partially due to the spread in mass.
We also measured the small separation \dnu{03} and tentatively identify a similar variation of 
the frequency separation ratio 
\dnu{03}/\Dnu\ as observed for \dnu{02}/\Dnu\ as a function of \Dnu.

\item The absolute width of the $l=1$ ridge for stars of higher luminosity is significantly narrower than 
for stars with low luminosity. This is the first quantitative confirmation of more efficient mode trapping as 
predicted by theory.
 
\item We presented a first estimate for the relation between pulsation amplitude and \numax\ for \kep\ 
red giants. We observed a distinct lack of low-amplitude stars for 
$\numax\gtrsim110\,\muHz$ and found that, for a given \numax, stars with lower \Dnu\ (and therefore 
higher mass, see Equation (10)) tend to show lower amplitudes than stars with higher \Dnu. 
Both observations can be explained with a mass dependence of the amplitude-\numax\ relation, and 
therefore suggest that the scaling relation for luminosity amplitudes of red giants needs to be revised.
\end{enumerate}

\acknowledgments
The authors gratefully acknowledge the \kep\ Science Team and all those who have contributed to the 
\kep\ mission for their tireless efforts which have made these results possible. 
We are also thankful to A. Miglio for his kind help with the stellar population synthesis and to 
our anonymous referee for his/her helpful comments. Funding for the 
\kep\ Mission is provided by NASA's Science Mission Directorate. DH acknowledges support by the 
Astronomical Society 
of Australia (ASA). DS and TRB acknowledge support by the Australian Research Council. The National Center for 
Atmospheric Research is a federally funded research and development center sponsored by the U.S. 
National Science Foundation. SH, YPE and WJC acknowledge support by the UK Science and Technology 
Facilities Council.

\bibliographystyle{apj}
\bibliography{references}

\end{document}